\documentclass[aps, prx,twocolumn,longbibliography,superscriptaddress,amsmath,amssymb,floatfix]{revtex4-2}

\usepackage{amssymb}
\usepackage{graphicx}
\usepackage{amsmath}
\usepackage{color}
\usepackage{mathrsfs}
\usepackage{float}
\usepackage{indentfirst}
\usepackage{mathrsfs}
\usepackage{float}
\usepackage{indentfirst}
\usepackage{textcomp}
\usepackage{comment}
\usepackage{mathtools}
\usepackage{natbib,hyperref}
\usepackage{soul}

\begin{document}
\title{Dismagicker: Unitary Gate for Non-Stabilizerness Reduction}
%%%%%%%%%%%%

%%%%%%%%%%%%
\author{Jiale Huang} \altaffiliation{These authors contributed equally to this work.}
\affiliation{Key Laboratory of Artificial Structures and Quantum Control (Ministry of Education),  School of Physics and Astronomy, Shanghai Jiao Tong University, Shanghai 200240, China}

\author{Rongyi Lv} \altaffiliation{These authors contributed equally to this work.}
\affiliation{Key Laboratory of Artificial Structures and Quantum Control (Ministry of Education),  School of Physics and Astronomy, Shanghai Jiao Tong University, Shanghai 200240, China}

\author{Xiangjian Qian} \altaffiliation{These authors contributed equally to this work.}
\affiliation{Key Laboratory of Artificial Structures and Quantum Control (Ministry of Education),  School of Physics and Astronomy, Shanghai Jiao Tong University, Shanghai 200240, China}
\affiliation{Tsung-Dao Lee Institute, Shanghai Jiao Tong University, Shanghai 200240, China}

\author{Mingpu Qin} \thanks{qinmingpu@sjtu.edu.cn}
\affiliation{Key Laboratory of Artificial Structures and Quantum Control (Ministry of Education),  School of Physics and Astronomy, Shanghai Jiao Tong University, Shanghai 200240, China}

\affiliation{Hefei National Laboratory, Hefei 230088, China}

\date{\today}

%%%%%%%%%%%%

%%%%%%%%%%%%
\begin{abstract}

We introduce the notion of \emph{dismagicker}: non-Clifford unitary gate designed to reduce the non-stabilizerness (also called magic) of quantum many-body states. Although both entanglement and non-stabilizerness are fundamental quantum resources, they require distinct control strategies. While disentanglers (unitary operations that lower entanglement) are well-established in tensor network methods, analogous concept for non-stabilizerness suppression has been largely missing. In this work, we define dismagicker as non-Clifford unitary operation that actively suppresses non-stabilizerness, steering states toward classically simulatable stabilizer states. We develop optimization method for constructing dismagickers within the Matrix Product States framework. Our numerical results show that the non-stabilizerness reduction procedure, when combined with entanglement reduction steps with Clifford circuits, significantly improves the accuracy for both classical simulation of many-body systems and quantum state preparation on quantum devices. Dismagicker enriches our toolkit for the manipulation of many-body states by unifying non-stabilizerness and entanglement reduction.
\end{abstract}
%%%%%%%%%%%%

%%%%%%%%%%%%
\maketitle
{\em Introduction --}
The study of quantum many-body systems represents a cornerstone of modern physics, driving progress in fields ranging from condensed matter to quantum information science. A central challenge within this domain is the identification and quantification of genuine quantum resources – properties that fundamentally distinguish quantum systems from their classical counterparts. Pinpointing these resources is not merely of theoretical interest; it is crucial for developing efficient classical simulation algorithms and for establishing provable quantum advantage in the burgeoning era of quantum computing \cite{Google_Nature}. Ultimately, understanding these resources is crucial to building effective quantum-classical hybrid methods that harness the strengths of both paradigms.

Entanglement and non-stabilizerness (also referred to as ``magic") constitute two fundamental, yet conceptually distinct quantum resources in many-body systems~\cite{RevModPhys.81.865,PhysRevA.71.022316,Veitch_2014,PhysRevX.6.021043,PhysRevLett.118.090501,howard2014contextuality, PRXQuantum.2.010345}. While both capture essential facets of the inherent ``quantumness" of a state, they probe fundamentally different aspects of its complexity. Entanglement quantifies the non-local correlations distributed between subsystems, serving as a cornerstone resource for quantum information processing \cite{RevModPhys.81.865}. However, the seminal Gottesman-Knill theorem \cite{gottesman1997stabilizer,PhysRevA.70.052328,PhysRevA.73.022334} imposes a critical constraint: entanglement alone is insufficient to reliably demarcate the boundary between computations that are efficiently classically simulatable and those possessing genuine, hard-to-simulate quantum complexity. This limitation is starkly illustrated by stabilizer states – states generated exclusively by Clifford circuits – which can exhibit extensive, volume-law entanglement entropy while remaining efficiently simulatable on classical computers~\cite{PhysRevX.7.031016}. Consequently, although entanglement is a necessary ingredient for quantum advantage, it is not a complete signature; additional resources are required to characterize the computational power beyond classical reach.

This gap is precisely filled by the concept of non-stabilizerness, which quantifies the deviation of a quantum state from the restricted set of stabilizer states (and thus, from efficient classical simulability under the Gottesman-Knill Theorem). Numerous resource-theoretic measures, such as the stabilizer extent~\cite{Bravyi2019simulationofquantum}, stabilizer fidelity~\cite{Bravyi2019simulationofquantum}, stabilizer rank~\cite{PhysRevX.6.021043,Bravyi2019simulationofquantum,PhysRevLett.116.250501}, robustness of magic~\cite{Heinrich2019robustnessofmagic}, Wigner negativity, mana~\cite{Veitch_2014,PhysRevLett.115.070501}, and stabilizer renyi entropy~\cite{PhysRevLett.128.050402} have been proposed to rigorously quantify this non-stabilizerness content, essentially capturing how far a given many-body state resides from the stabilizer polytope. Crucially, the relationship and interplay between entanglement and non-stabilizerness are profound and are actively being studied \cite{PhysRevB.110.045101,PhysRevA.109.L040401,qian2024augmentingdensitymatrixrenormalization,gxdn-zwrw}. Understanding how these two key resources co-evolve, constrain each other, or can be independently manipulated is essential for characterizing the complexity landscape of quantum many-body states and processes, forming a vital foundation for assessing quantum advantage and developing novel simulation methods.

Building upon the framework of entanglement as a resource, the concept of disentangler emerged: unitary operation specifically designed to reduce the entanglement of a quantum many-body state~\cite{PhysRevLett.99.220405,PhysRevB.111.035119}. Disentanglers have become indispensable building blocks in approaches like Tensor Network States~\cite{PhysRevLett.69.2863, PhysRevLett.75.3537, RevModPhys.93.045003, xiang2023density} (notably in the Multi-scale Entanglement Renormalization Ansatz, MERA~\cite{PhysRevLett.102.180406, PhysRevLett.99.220405}) and in the analysis of quantum circuit complexity. Their utility underscores the importance of understanding how quantum resources can be dynamically manipulated.

In this work, we introduce a conceptually analogous operation focused on the other key quantum resource: non-stabilizerness. We define dismagicker as deliberately engineered unitary gate designed to systematically reduce the non-stabilizerness of a quantum many-body state. Crucially, since Clifford circuits preserve non-stabilizerness, a dismagicker must inherently be a non-Clifford operation—thereby explicitly manipulating non-stabilizerness as a distinct quantum resource. By introducing and characterizing dismagickers, we establish a new theoretical and operational framework for the controlled dissipation of non-stabilizerness, mirroring the role of disentanglers in managing entanglement.

To demonstrate the practical utility of this new concept, we develop method that leverages dismagickers to reduce non-stabilizerness in quantum many-body simulations. We test the effectiveness of dismagicker across different many-body systems including random many-body states and Heisenberg model, showing that it can significantly improve the efficiency of quantum state preparation and simulation accuracy. This work opens pathways for novel quantum state engineering techniques, resource theory development, and new classical simulation strategies targeting the non-stabilizerness resource.

{\em non-stabilizerness and its measure --} Within the resource theory of quantum computation, non-stabilizerness, also known as ``magic", constitutes the fundamental resource necessary to surpass classical simulation. While a variety of quantifiers have been proposed, including stabilizer extent~\cite{Bravyi2019simulationofquantum}, stabilizer fidelity~\cite{Bravyi2019simulationofquantum}, stabilizer rank~\cite{PhysRevX.6.021043,Bravyi2019simulationofquantum,PhysRevLett.116.250501}, robustness of magic~\cite{Heinrich2019robustnessofmagic}, Wigner negativity and mana~\cite{Veitch_2014,PhysRevLett.115.070501}, the Stabilizer Rényi Entropy (SRE) has emerged as a effectively computable metric~\cite{PhysRevLett.128.050402, Haug2023stabilizerentropies, PhysRevLett.131.180401, PhysRevB.107.035148, PhysRevLett.133.010601}. The SRE of index $\alpha$ for a pure normalized state $|\psi\rangle$ is defined as
\begin{equation}
    M_\alpha(|\psi\rangle) = \frac{1}{1-\alpha} \log_2 \sum_{P\in\mathcal{P}_n} \frac{|\langle\psi|P|\psi\rangle|^{2\alpha}}{2^n},
\end{equation}
where $\mathcal{P}_n$ denotes the $n$-qubit Pauli string. In this work, we focus on the $\alpha=2$ case, $M_2(|\psi\rangle)$, which serves as our primary cost function for quantifying non-stabilizerness.

The motivation for adopting $M_2$ extends beyond its practical computability; it is deeply rooted in its connection to the stabilizer fidelity. Naturally, finding the nearest stabilizer state to a target $|\psi\rangle$ requires maximizing the stabilizer fidelity, $F_{\text{stab}}(|\psi\rangle) = \max_{|\phi\rangle \in \text{STAB}} |\langle \psi | \phi \rangle|^2$~\cite{Bravyi2019simulationofquantum}. Since direct optimization this overlap is computationally intractable, the SRE offers a rigorous and efficient surrogate. Crucially, it bounds the logarithmic fidelity via the relation $F_{\text{stab}} \ge 2 e^{-M_2} - 1$ \cite{Haug2025probingquantum,PhysRevLett.132.240602}. This inequality  guarantees that variationally minimizing $M_2$ to a sufficiently small value forces the state to maximize its overlap with a stabilizer state. Consequently, suppressing the SRE effectively evolves the target state toward the nearest stabilizer state.

{\em Dismagicker as unitary gate for non-stabilizerness reduction --}
To actively manipulate and compress this quantum resource, we introduce the concept of \emph{dismagicker}, a specialized unitary operation, $U_{\mathcal{M}}$, designed to deplete the non-stabilizerness of a target state $|\psi\rangle$. In this work, we formally seek the unitary that minimizes the non-stabilizerness, formulated as $\min_{U_{\mathcal{M}}} M_2(U_{\mathcal{M}}|\psi\rangle)$. Because the unitary group provides a continuous covering of the Hilbert space, there exist infinitely many continuous trajectories connecting any arbitrary state $|\psi\rangle$ to a stabilizer state, allowing us to systematically drive complex quantum states toward these classically simulatable states.

\begin{figure}
    \centering
    \includegraphics[width=85mm]{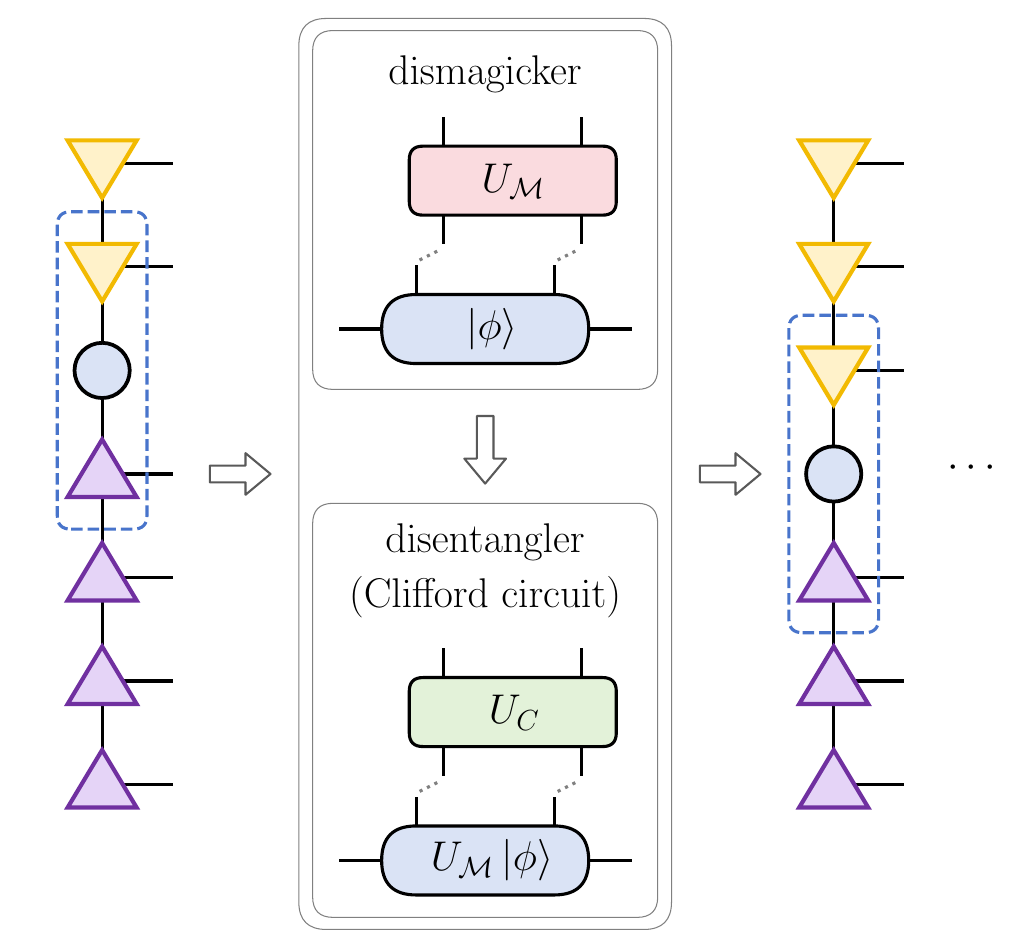}
    \caption{Schematic of the interleaved optimization flow of dismagicker and Clifford disentangler on a MPS. At each step of the sweep, a local two-site tensor $|\phi\rangle_{k,k+1}$  is targeted. We first apply a parameterized dismagicker gate $U_{\mathcal{M}}$  to minimize the non-stabilizerness. Immediately following this, a two-site Clifford disentangler $U_C$ is applied to the intermediate state $U_{\mathcal{M}}|\phi\rangle$  to actively suppress the entanglement entropy without altering the non-stabilizerness. Finally, a Singular Value Decomposition (SVD) is performed to update the local tensors, after which the algorithm moves to the next pair of sites to continue the sweep.}
    \label{fig:UCUM}
\end{figure}

However, minimizing non-stabilizerness is a fundamentally distinct objective from minimizing entanglement. The independence of these two resources is demonstrated by contrasting the extensively non-stabilizerness but strictly unentangled product state $|T\rangle^{\otimes n}$ (with $T=\frac{1}{\sqrt{2}}(|0\rangle + e^{i\pi/4}|1\rangle)$) with the maximally entangled yet entirely magic-free GHZ state (GHZ$=\frac{1}{\sqrt{2}}(|0\rangle^{\otimes n} + |1\rangle^{\otimes n})$). Consequently, applying a generic dismagicker $U_{\mathcal{M}}$ solely to minimize the $M_2$ typically doesn't reduce the Entanglement Entropy (EE). This reveals a tension: while optimizing a quantum state purely for stabilizerness successfully reduces $M_2(U|\psi\rangle)$, it does not necessarily yield a state that is easier to simulate classically in the framework of Tensor Network States.

To mitigate this issue, we propose a synergistic approach by integrating dismagicker with Clifford disentanglers ($U_C$) within Matrix Product State (MPS) framework, as schematically illustrated in Fig.~\ref{fig:UCUM}. Because Clifford circuits are strictly magic-preserving and leave the $M_2$ invariant, an optimal Clifford disentangler can be applied to actively suppress the EE while fully retaining the $M_2$ reduction. Crucially, a joint optimization that interleaves these two operations ($U_C U_{\mathcal{M}}$) structurally outperforms a naive sequential dismagicker application, as we will show below.

{\em Test results --} We first benchmark the effectiveness of dismagicker in suppressing non-stabilizerness using random 6-qubit states. To ensure the test states exhibit both substantial entanglement and substantial non-stabilizerness, we prepare them in two stages. We first generate a highly entangled stabilizer state by applying a depth-6 circuit consisted of random two-qubit Clifford circuits to a product state. We then inject non-stabilizerness by applying three additional layers of Haar-random two-qubit unitary gates to this intermediate state.
\begin{figure}
    \centering
    \includegraphics[width=85mm]{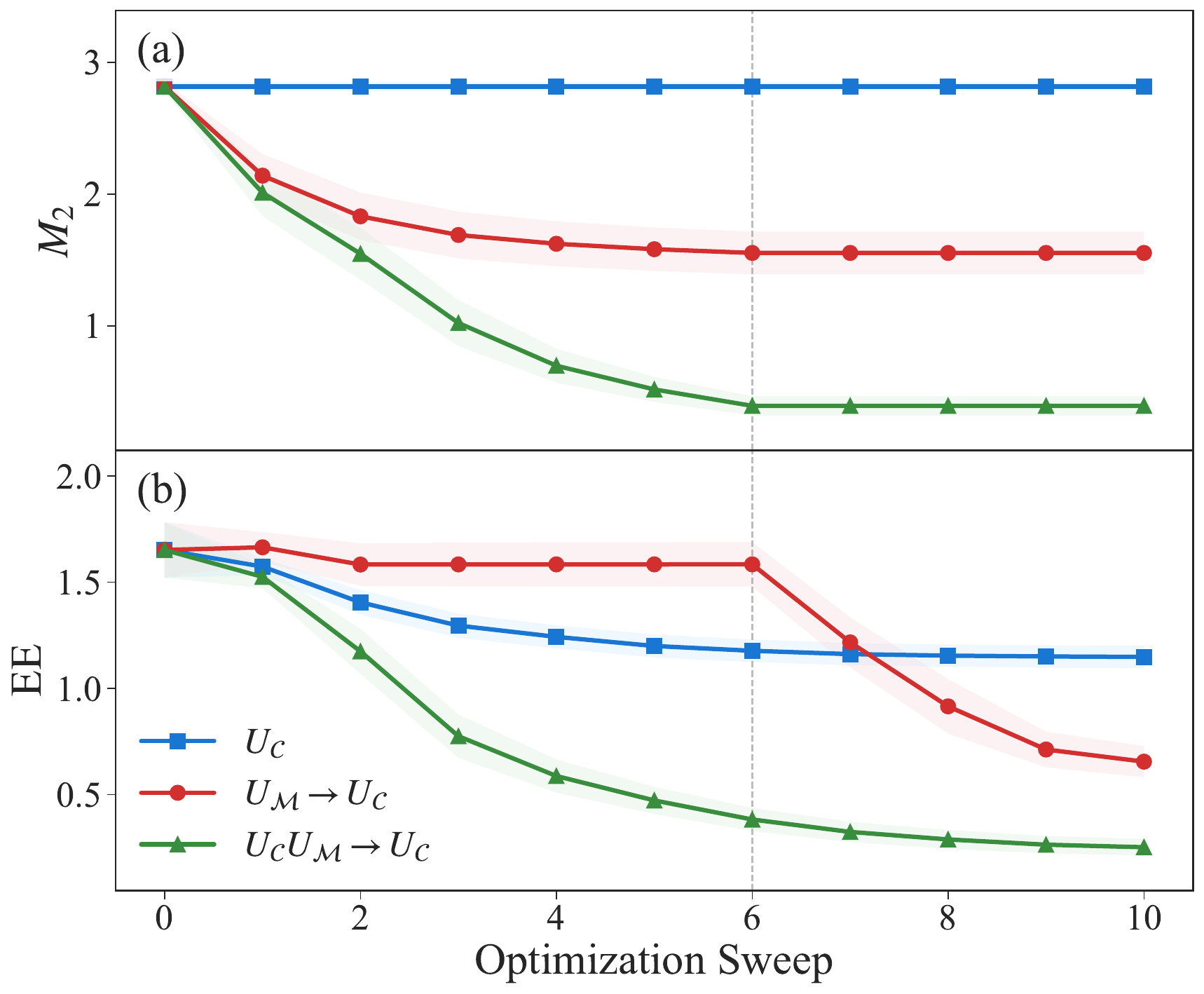}
    \caption{The reduction of $M_2$ (a) and EE (b) with different strategies. The protocol transitions from a dismagicker phase (sweeps 1–6) to a Disentangler phase (sweeps 6–10) for the red and green results. Joint optimization ($U_C U_{\mathcal{M}} \rightarrow U_C$, green) achieves significantly deeper simultaneous suppression of both resources compared to sequential ($U_{\mathcal{M}} \rightarrow U_C$, red) or Clifford-only ($U_C$, blue) strategies. Results are averaged over $1000$ random initial $N=6$ states. Highly entangled stabilizer states are first generalized by applying a depth-6 circuit consisted of random two-qubit Clifford circuits to a product state. Then non-stabilizerness is injected by applying three additional layers of Haar-random two-qubit unitary gates to this intermediate state. Standard deviations are represented by the shadow.}
    \label{fig:N6}
\end{figure}

We compare three strategies in this work. In the first strategy, we apply two-qubit Clifford circuits at each step to suppress entanglement alone, since Clifford circuits preserve non-stabilizerness. We obtain the optimal circuit by enumerating all possible two-qubit Clifford circuits, as the search space is small. In the second strategy, we employ dismagicker to suppress non-stabilizerness without accounting for entanglement. Each local two-qubit dismagicker operation is represented as a parameterized two-qubit gate $e^{iV(\theta)}$ , defined by a 16-parameter Hermitian generator $V(\theta)$. The optimal local unitary is then determined by variationally minimizing the exact $M_2$ using the gradient-free Nelder-Mead algorithm. After a few step of non-stabilizerness suppression with dismagickers, we then apply the first strategy to reduce the entanglement entropy using Clifford circuit. In the third strategy, we first identify the optimal two-qubit dismagicker and then apply a two-qubit Clifford circuit to reduce entanglement in each step. After a few sweeps of this procedure, pure Clifford circuits are also employed to reduce the entanglement entropy.

Fig.~\ref{fig:N6} tracks the evolution of $M_2$ and EE by averaging $1000$ random state realizations. As expected, pure Clifford operations (blue in Fig.~\ref{fig:N6}) preserve the initial non-stabilizerness, leaving the $M_2$ strictly invariant. But the entanglement entropy is reduced \cite{gxdn-zwrw}. In contrast, the sequential application of the dismagicker (red in Fig.~\ref{fig:N6}) forces a rapid $M_2$ reduction, directly demonstrating that actively depleting non-stabilizerness is practically achievable. However, optimizing non-stabilizerness in isolation typically conflicts with EE minimization. Consequently, this naive approach plateaus at a suboptimal value and sustains highly entanglement during the dismagicker phase (sweeps 1–6). Nevertheless, this dismagicks still enables the subsequent Disentangler phase to achieve a lower final entanglement than the pure Clifford  strategy as shown in Fig.~\ref{fig:N6} (b).

In the third strategy, by appending a Clifford disentangler $U_C$ immediately after each $U_{\mathcal{M}}$, the algorithm actively suppresses the EE while dynamically assisting the dismagicker in bypassing $M_2$ traps (green in Fig.~\ref{fig:N6}). This synergistic approach achieves a substantially deeper, simultaneous reduction of both resources. After the subsequent disentangler phase with Clifford circuits (sweeps 6–10), the state is seamlessly driven into a highly compressible, low-entanglement regime. Ultimately, this confirms that the dismagicker can effectively prepare substantially simpler initial states for quantum computation, drastically reducing the non-Clifford resource overhead required for state preparation.

We remark, however, that the final $M_2$ does not strictly vanish to zero in our test results. Since the non-stabilizerness in the initial states is injected via exactly three layers of random unitaries, a theoretically perfect inversion could completely annihilate the non-stabilizerness within three sweeps. This persistent residual non-stabilizerness highlights the inherent difficulty of fully depleting non-stabilizerness. Overcoming this bottleneck by exploring alternative optimization strategies or different non-stabilizerness measure remains an open challenge for future development.

\begin{figure}
    \centering
    \includegraphics[width=85mm]{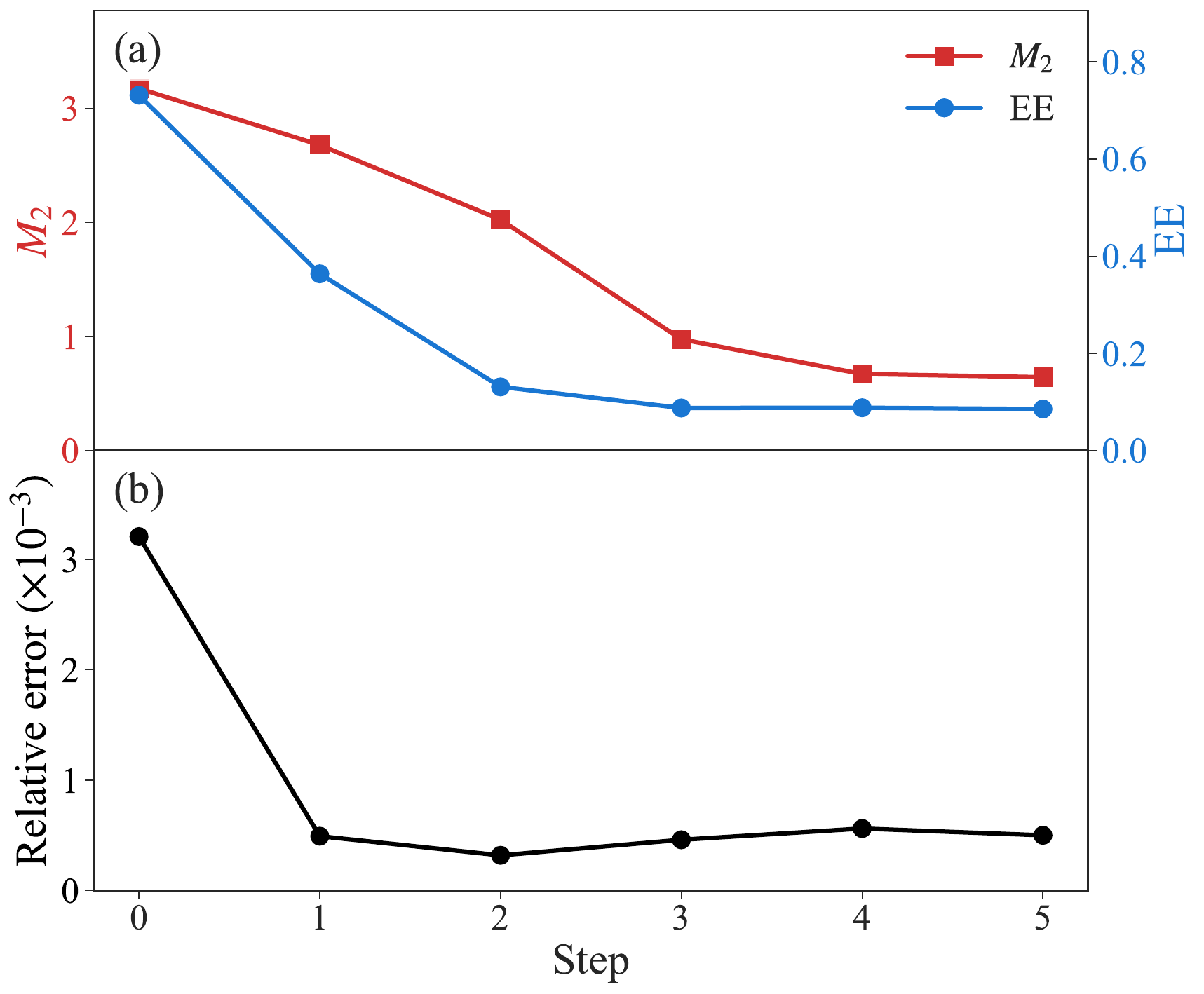}
    \caption{Result from the joint optimization of dismagicker and Clifford disentangler of the 1D Heisenberg chain ground state with size $L=20$. The initial state is an MPS with bond dimension $D=4$ from DMRG. (a) Evolution of $M_2$ and EE versus optimization sweeps. The dismagicker $U_{\mathcal{M}}$ is selected by sampling $200$ random Clifford+$R_z(\theta)$ gate combinations, with $M_2$ evaluated via sampling from $10^4$ shots \cite{PhysRevLett.131.180401}. (b) Relative error $|E - E_\text{exact}|/|E_\text{exact}|$ of the ground-state energy. At each step, a DMRG calculation restricted to $D=4$ is performed on the effectively transformed Hamiltonian, demonstrating rapid convergence to a much higher accuracy than the initial DMRG result.}
    \label{fig:L20}
\end{figure}

We also demonstrate the practical utility of our protocol in many-body physics by applying it to the ground state of a 1D Heisenberg chain with size $L=20$. The Hamiltonian of Heisenberg model is $H = \sum_i^{L-1} \hat{S}_i\hat{S}_{i+1} $, where $\hat{S}_i = (S^x_i, S^y_i, S^z_i)$ is the spin-1/2 operator for site $i$. The initial state is represented as a MPS with a bond dimension $D=4$ from Density Matrix Renormalization Group (DMRG) \cite{PhysRevLett.69.2863}. To implement the local updates efficiently, we employ a discrete sampling strategy: the optimal $U_{\mathcal{M}}$ is selected from 200 random Clifford and $R_z(\theta)$ gate combinations, while the SRE is efficiently evaluated via MPS-based perfect sampling with $10^4$ shots \cite{PhysRevLett.131.180401}. Consistent with the $N=6$ case for random states, the joint sweeps with dismagicker and Clifford disentangler successfully suppress both the $M_2$ and EE as shown in Fig.~\ref{fig:L20} (a). 

Fig.~\ref{fig:L20} (b) displays the relative error of the ground-state energy (defined as $|E - E_\text{exact}|/|E_\text{exact}|$), obtained by performing a fresh DMRG calculation on the effectively transformed Hamiltonian, restricted to the same bond dimension $D=4$. As the optimization actively depletes the inherent non-stabilizerness and entanglement, the relative energy error drops precipitously and converges to a highly stable regime. This provides compelling numerical evidence that the dismagicker successfully rotates the physical problem into a computationally simpler basis, directly enhancing the precision of tensor network methods.

We also attempted to integrate non-stabilizerness reduction directly into the DMRG framework, drawing on the approach used in Clifford circuits Augmented Matrix Product States (CAMPS) \cite{qian2024augmentingdensitymatrixrenormalization}, where unitary optimization is embedded within the ground-state search. However, the resulting suppression of $M_2$ is not as pronounced as shown in Fig. \ref{fig:L20}. We find that incorporating the dismagicker optimization directly into the DMRG sweep introduces a fundamental tension between competing quantum resources: DMRG truncates the bond dimension based on the entanglement spectrum, whereas the dismagicker actively targets non-stabilizerness. Critically, these entanglement-driven truncations disrupt the non-stabilizerness structure of the wavefunction, thereby compromising the optimization of the non-stabilizerness. In future work, we will explore ways to resolve this conflict.

{\em Discussion --}
Although dismagicker is proposed to target non-stabilizerness, its action inherently affects entanglement as well. This coupling arises because dismagickers are defined up to Clifford circuits, which leave non-stabilizerness invariant but can be used to tune entanglement properties. Consequently, the dismagicker framework provides a unified approach for suppressing both key quantum resources. In practical applications, such as in DMRG, the conjugation of general dismagickers complicates the Hamiltonian during optimization. Balancing the resulting increase in computational cost against the benefits of reduced non-stabilizerness and entanglement requires careful design and is expected to be system-dependent. However, this overhead can be avoided by constraining the dismagicker to a specific class of operations, such as Matchgates in the study of fermionic systems \cite{2026-0062}. 

{\em Conclusion and Perspective --}
In this work, we introduce the notion of dismagicker, non-Clifford unitary gate that reduces non-stabilizerness in quantum many-body states—serving as an analogue to disentangler. We present a practical method for optimizing dismagicker gates which reduce the non-stabilizerness of a given state within the MPS framework, using SRE to quantify non-stabilizerness. When integrated with entanglement reduction with Clifford circuits, this approach significantly improves accuracy for both classical simulation of many-body systems and quantum state preparation on quantum devices.

Several avenues for future work remain: adopting other non-stabilizerness measures to further test the efficiency of our method, developing (local) proxies for non-stabilizerness measures to reduce computational cost, developing more efficient and effective optimization strategies, and extending the framework beyond matrix product states to other computational paradigms. Our test in this work has focused on applying dismagickers to a fixed many-body state. A natural and important extension is the application to dynamical problems, such as time evolution, where the growth of both entanglement and non-stabilizerness \cite{PhysRevLett.134.150404,PhysRevLett.134.150403} fundamentally limits classical simulatability, offering a critical testbed for our technique. Collectively, dismagickers establish a useful tool for advancing the accuracy and efficiency of quantum many-body simulation and quantum circuits analysis.

\textbf{Acknowledgments:} The computation in this paper were run on the Siyuan-1 cluster supported by the Center for High Performance Computing at Shanghai Jiao Tong University. MQ acknowledges the support from the National Natural Science Foundation of China (Grant No. 1252240 and No. 12274290), the Innovation Program for Quantum Science and Technology (2021ZD0301902), and the National Key Research and Development Program of MOST of China (2022YFA1405400).
%%%%%%%%%%%%

%%%%%%%%%%%%Refs
\bibliography{main}
%%%%%%%%%%%%

\end{document}